\documentclass[final,letter]{aipproc}
\layoutstyle{8x11double}

\newcommand{\sun}{\hbox{$\odot$}}

\begin{document}

\title{Timing Analysis of Light Curves in the Tartarus Active Galactic
Nuclei Database}

\author{Paul  M.    O'Neill}{  address={Astrophysics  Group,  Imperial
College London,  Blackett Laboratory, Prince Consort  Road, London SW7
2AW, United Kingdom} }

\author{Kirpal Nandra}{  address={Astrophysics Group, Imperial College
  London, Blackett  Laboratory, Prince  Consort Road, London  SW7 2AW,
  United Kingdom} }

\author{Iossif   E.   Papadakis}{   address={Department   of  Physics,
University  of Crete,  71 003,  Heraklion,  Greece} ,altaddress={IESL,
FORTH-Hellas, 71 110, Heraklion,  Crete, Greece} }

\author{T.    Jane  Turner}{   address={Laboratory  for   High  Energy
Astrophysics, Code  660, NASA Goddard Space  Flight Center, Greenbelt,
MD 20771} ,altaddress={University  of Maryland, Baltimore County, 1000
Hilltop Circle, Baltimore, MD 21250} }

\begin{abstract}

The Tartarus database contains products for 529 \emph{ASCA}
observations of active galactic nuclei.  We have been updating
Tartarus to include observing sequences conducted after 1999
January. The revised database will contain products for 375 objects,
with a total of 614 observing sequences. We have begun a systematic
timing analysis of the Tartarus light curves. We present here some
preliminary results of an investigation into the relation between
excess variance and black-hole mass. Having optimised our analysis
to minimize the scatter in the variance measurements, we find that
the narrow-line active galactic nuclei follow roughly the same
relation as the broad-line objects.

\end{abstract}

\maketitle


\section{X-RAY VARIABILITY IN ACTIVE GALACTIC NUCLEI}

The  existence  of  variability  in  the X-ray  emission  from  active
galactic  nuclei  (AGNs),  on   time-scales  of  days  to  years,  was
established  roughly  three-decades  ago  \citep[e.g.][and  references
therein]{mwp81}.  Variations on time-scales shorter than this are also
common.  The launch of \emph{EXOSAT},  with its long orbit, meant that
well-sampled X-ray  light curves could be obtained  for time-scales of
minues to days.  The  power-spectra generated from \emph{EXOSAT} light
curves    could    be    described    by   an    unbroken    power-law
\citep[e.g.][]{lp93,gml93}.  Archival observations spanning weeks
to months showed that, for two objects, the power-law flattened at low
frequencies   \citep{m88,pm95}.    More   recently,  \emph{RXTE}   and
\emph{XMM-Newton}  observations have shown  that the  power-spectra of
AGN light curves exhibit broken power-law shapes very similar to those
seen       in      galactic       black-hole       X-ray      binaries
\citep[e.g.][]{mpu03,mev03}.

As well  as power-spectral density, the \emph{excess  variance} can be
used       to      characterise       the       X-ray      variability
\citep[e.g.][]{ngm97}. Excess  variance is defined as  the variance in
the  light curve  not due  to statistical  fluctutations.   The window
function  affects   the  excess  variance   less  than  it   does  the
power-spectral  density.    Therefore,  the  excess   variance  is  an
especially  useful  quantity  for  the  description  of  light  curves
containing many gaps.

An anti-correlation  was found between excess  variance and luminosity
for  a sample  of AGNs  observed by  the \emph{Advanced  Satellite for
Cosmology and  Astrophysics} (\emph{ASCA}) \citep{ngm97}.   Later work
using \emph{ASCA} data  found that, for a given  luminosity, the X-ray
light curves of narrow-line Seyfert~1 (NLS1) galaxies exhibit a larger
excess   variance   than    broad-line   Seyfert~1   (BLS1)   galaxies
\citep{tgn99,l99}.   This result  can  be explained  if,  for a  given
luminosity, the NLS1s have a  lower black-hole mass than the BLS1s.  A
lower mass results in faster variations,  and this is seen as a larger
excess variance.  Alternatively,  other properties of the variability,
not just the  time-scale, might differ between NLS1s  and BLS1s. These
alternatives  can be tested  through examining  the dependence  of the
excess variance on black-hole mass.

The  relationship between  excess  variance and  black-hole mass,  for
Seyfert 1 galaxies and  quasi-stellar objects (QSOs), has been studied
using  \emph{ASCA}  data  (time-scale $\sim  1$~d)  \citep{ly01,bz03}.
These studies revealed an  anti-correlation between mass and variance.
Moreover, the  narrow-line (NL AGNs) and broad-line  (BL AGNs) objects
followed the same  relation.  This relation could be  explained if the
power-spectrum   is   an   unbroken   power-law  [$P(\nu   )   \propto
\nu^{-\alpha}$] with $\alpha \sim 2$ \citep{ly01}.

An  investigation  using \emph{RXTE}  data  (time-scale $\sim  300$~d)
presents  a different picture  \citep{p03}.  The  BLS1s, and  the NLS1
MCG--6-30-15, follow a variance--mass  relation that is consistent with
a universal power-spectral shape.  This `average' spectrum comprises a
twice-broken power-law  which breaks from $\alpha  = 0$ to  $1$ at the
so-called `low-frequency break',  and from $\alpha = 1$  to $2$ at the
`high-frequency   break'.   The   break   frequencies  are   inversely
proportional to  black-hole mass, and the  amplitude, when represented
in  power$\times$frequency space,  is  constant.  This  power-spectral
shape  resembles that  seen in  the black-hole  binary Cyg~X-1  in the
low/hard  state.  The other  NLS1 in  the sample,  NGC~4051, \emph{did
not} follow the  above relation.  The excess variance  of NGC~4051 was
consistent with a singly-broken  power-law, breaking from $\alpha = 1$
to $2$,  with a  break frequency  that is $20$  times higher  than for
BLS1s.  This shape is similar to that seen in Cyg~X-1 in the high/soft
state.   A  power-spectral  analysis  of NGC~4051,  using  \emph{RXTE}
(time-scale $\sim 6.5$~years)  and \emph{XMM-Newton} (time-scale $\sim
100$~ks) data,  confirms that NGC~4051  is indeed an analogue  to high
state  binaries \citep{mpu03}.  Moreover,  in plots  of high-frequency
break  versus   black-hole  mass,  the   NLS1s  were  found   to  have
systematically    higher   break    frequencies    than   the    BLS1s
\citep{p03,mpu03}.

The excess  variance versus luminosity relation for  the BLS1 galaxies
can be explained as a consequence of the variance-mass relation if the
BLS1s  radiate  at  $\sim  13$~per~cent of  the  Eddington  luminosity
\citep{p03}.   The  location of  NGC~4051  in the  variance-luminosity
plane is  consistent with that object radiating  at $\sim 10$~per~cent
Eddington \citep{p03}.

These results from \emph{RXTE}  and \emph{XMM-Newton} suggest that the
power-spectral properties of AGN  depend on black-hole mass \emph{and}
at least one other parameter.  The dependence with mass is perhaps the
least  surprising,   since  we  expect   the  size  of   the  emission
region---and,  therefore, the  variability time-scale---to  scale with
mass.   A useful  step  forward, then,  is  to account  for this  mass
dependence and investigate how  the X-ray variability depends on other
parameters.  This is best  achieved through maximising the sample size
and minimising observational scatter.

\section{TARTARUS: A DATABASE OF ASCA OBSERVATIONS OF AGN}

The \emph{ASCA} satellite was launched  into a 96~minute orbit on 1993
February    20    and   ceased    observing    on    2000   July    14
\citep{tih94}\footnote{see                                         also
\url{http://heasarc.gsfc.nasa.gov/docs/asca/ascagof.html}}.         The
satellite  comprised  four  X-ray  telescopes and  four  corresponding
focal-plane  instruments \citep{sjs95}.   There  were two  solid-state
imaging spectrometers (SIS0 and  SIS1) \citep{g95} and two gas imaging
spectrometers  (GIS2 and  GIS3) \citep{oef96}.   The  effective energy
ranges  were   0.5--10~keV  and  0.7--10~keV  for  the   SIS  and  GIS
instruments, respectively.

The   Tartarus  database   contains  products   for   all  \emph{ASCA}
observations  with  targets  designated  as  AGN  \citep{tnt01}.   The
products comprise SIS and GIS  event files, images, spectra, and light
curves.               Version~2              is              presently
available\footnote{\url{http//:tartarus.gsfc.nasa.gov}}   at   Goddard
Space   Flight  Centre   (GSFC),   and  consists   of  529   observing
sequences. We are updating and  improving the database to include data
collected after 1999 January. The revised database will consist of 614
sequences with a total of 375 objects.  Version~3 of Tartarus is to be
made available at GSFC and Imperial College London.

We are conducting a systematic  timing analysis of the light curves in
the  Tartarus database.   Tartarus is  well suited  to such  a project
because it contains  data for a large sample  of objects.  The overall
aim of the  project is to uncover any  correlations that exist between
the properties of the X-ray variability and other parameters.

We have  begun the  project by investigating  those AGN for  which the
black-hole mass has been  estimated.  We present here some preliminary
results  of an  investigation between  excess variance  and black-hole
mass. Our work is similar  to that conducted using \emph{RXTE} data as
described in the previous section.

\section{ANALYSIS}

\begin{table}
\begin{tabular}{lcccc}
\hline \tablehead{1}{l}{t}{Object\\Name} & \tablehead{1}{c}{t}{Type} &
\tablehead{1}{c}{t}{Mass\\($10^{7}$ M$_{\sun}$)}
& \tablehead{1}{c}{t}{Ref.}  \\

\hline

{\bf Variable} \\

NGC~4051 & NL AGN & $0.05^{+0.06}_{-0.03}$ & \citep{sun03} \\

MRK~766 & NL AGN & 0.0832 & \citep{bz03} \\

AKN~564 & NL AGN & 0.115 & \citep{bz03} \\

MCG$-$6-30-15 & NL AGN & 0.155 & \citep{bz03} \\

MRK~1040 & BL AGN & 0.229 & \citep{bz03} \\

NGC~4593 & BL AGN & $0.66\pm0.52$ & \citep{opd03} \\

NGC~7469 & BL AGN & $0.75^{+0.74}_{-0.75}$ & \citep{ksn00} \\

MRK~110 & NL AGN & $0.77^{+0.28}_{-0.29}$ & \citep{ksn00} \\

NGC~3783 & BL AGN & $0.87\pm0.11$ & \citep{op02} \\

MRK~290 & BL AGN & 1.12 & \citep{bz03} \\

NGC~4151 & BL AGN & $1.20^{+0.83}_{-0.70}$ & \citep{ksn00} \\

TON~S180 & NL AGN & 1.23 & \citep{bz03} \\

NGC~3516 & BL AGN & $1.68\pm0.33$ & \citep{opd03} \\

NGC~3227 & BL AGN & $3.6\pm1.4$ & \citep{opd03} \\

MRK~509 & BL AGN & $9.2\pm1.1$ & \citep{ksn00} \\

NGC~5548 & BL AGN & $9.4^{+1.7}_{-1.4}$ & \citep{ksn00} \\

\hline

{\bf Non-Variable} \\

MRK~335 & NL AGN & $0.38^{+0.14}_{-0.10}$ & \citep{ksn00} \\

IC~4329A & BL AGN & $0.7^{+1.8}_{-1.6}$ & \citep{ksn00} \\

PG~0026$+$12 & NL AGN & $2.66^{+0.49}_{-0.55}$ & \citep{ksn00} \\

MRK~279 & BL AGN & 4.17 & \citep{bz03} \\

F~9 & BL AGN & $8.3^{+2.5}_{-4.3}$ & \citep{ksn00} \\

MRK~841 & BL AGN & 14.1 & \citep{bz03} \\

PG~1116$+$215 & BL AGN & 16.2 & \citep{wu02} \\

PG~0804$+$761 & BL AGN & $16.3^{+1.6}_{-1.5}$ & \citep{ksn00} \\

AKN~120 & BL AGN & $18.7^{+4.0}_{-4.4}$ & \citep{ksn00} \\

MCG$-$2-58-22 & BL AGN & 34.7 & \citep{bz03} \\

MRK~1383 & BL AGN & $37^{+13}_{-16}$ & \citep{ksn00} \\

IRAS~F04250$-$5718 & BL AGN & 38.0 & \citep{bz03} \\

HE~1029-1401 & BL AGN & 120 & \citep{wu02} \\

\hline

\end{tabular}

\caption{Name,  type and  black-hole mass  for each  object  having at
least  1   valid  Tartarus   light  curve.   The   objects  exhibiting
significant variability are listed first.}

\label{tab:objinfo}
\end{table}

We restricted our  analysis to narrow- and broad-line  AGN. Our sample
contains Seyfert~1 galaxies and QSOs.  We do not differentiate between
Seyfert galaxies and QSOs, but we shall simply refer to each object as
either a  narrow- or  broad-line AGN \citep{bz03}.   We used  the `rms
spectrum'   \citep{pwb98}  reverberation-mapping   mass   estimate  if
available,  otherwise we used  the mass  estimate determined  from the
optical  luminosity \citep[e.g][]{wu02}.  The  objects in  our sample,
along    with    their    black-hole    masses,    are    listed    in
Table~\ref{tab:objinfo}.   The mass  measurements with  no  error bars
were estimated  via the optical luminosity.  The  uncertainty in these
measurements   is   about    0.5~dex   \citep{bz03}.    Objects   with
$\mathrm{H}\beta  <  2000$~km  s$^{-1}$  were classified  as  NL  AGNs
\cite{tgn99,bz03,bg92,md01}.

The  variance  in a  light  curve  is equal  to  the  integral of  the
power-spectral density  between the frequency  limits set by  the time
resolution and  duration of  the light curve.   Therefore, the  use of
different binnings and durations  per object will introduce artificial
scatter.  We wish to minimize this scatter.  Therefore, we used a time
resolution of  256~s for all light  curves, and restricted  them to be
between 30 and 40~ks in duration.

A sample of measurements of the excess variance of a red-noise process
will possess  a large scatter  about their mean \citep[][see  also the
following  section]{vew03}.  This  scatter is,  of course,  reduced by
averaging over  many realizations.  Therefore, in  our analysis, light
curves longer  than 40~ks were subdivided into  many shorter segments.
This allowed  us to maximize  the number of variance  measurements per
object.

We extracted a  2--10~keV combined SIS0 and SIS1  light curve for each
object. We  only used light  curves comprising at least  20~bins, with
each bin  having no fewer  than 20~counts.  Some light  curves, having
being  initially  extracted, contained  some  bins  having fewer  than
20~counts. In  these cases, we excised  bins from the  light curve, on
the   basis  of  fractional   exposure,  so   that  only   valid  bins
remained. The invalid bins could  thus be removed from the light curve
without introducing bias.

There were 29~objects  for which we had at least  a single valid light
curve segment. These objects are listed in Table~\ref{tab:objinfo}. We
determined  the  $\chi^{2}$  corresponding  to  the  hypothesis  of  a
constant counting  rate for each segment.   We then summed  all of the
$\chi^{2}$s and degrees-of-freedom for each object, and thereby tested
whether that object exhibited variability.  We detected variability in
16  objects  at the  95~per~cent  confidence  significance level  (see
Table~\ref{tab:objinfo}).  We  are yet  to determine the  upper limits
for those objects in which variability was not detected.

For the 8  objects having at least 5  measurements of excess variance,
the uncertainty  in the mean  variance was measured from  the observed
scatter in  those measurements. The standard deviation  of the scatter
about the  mean was found  to be $\sim  60$~per~cent of the  mean.  We
used this  relation to assign  uncertainties for the 8  objects having
fewer than 5 variance measurements.

\begin{figure}
  \rotatebox{270}{\includegraphics[height=1\columnwidth]{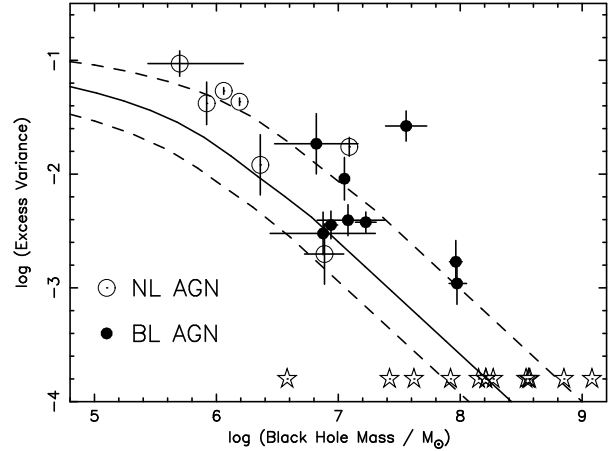}}
  \caption{Log$_{10}$   of  excess   variance  versus   log$_{10}$  of
black-hole  mass.   Open  and  filled circles  represent  NL~AGNs  and
BL~AGNs, respectively.  The stars indicate the masses of those objects
in  which variability  was not  detected.   The solid  line shows  the
expected variance  assuming a twice-broken  power law, and  the dashed
lines show the range in the expected value. }
  \label{fig:xrt2003plot}
\end{figure}

\subsection{RESULTS, DISCUSSION AND FUTURE DIRECTIONS}

A plot of  log$_{10}$ of the excess variance  versus log$_{10}$ of the
black-hole mass  is presented in  Fig.~\ref{fig:xrt2003plot}. The open
and  filled circles  indicate the  NL~AGNs and  BL~AGNs, respectively.
The stars indicate  the masses of the 13 objects in  we did not detect
variability.

The  solid  line  in  Fig.~\ref{fig:xrt2003plot}  shows  the  expected
variance assuming a twice-broken  power-law.  We determined this using
the relation found from \emph{RXTE} data for BLS1s \citep{p03}, except
that we also  accounted for red-noise leak. The  dashed lines indicate
the rough range of the expected  variance based on the errors from the
\emph{RXTE} analysis.

The  NL~AGNs  and BL~AGNs  follow  roughly  the  same relation.   This
conclusion was also reached with previous analyses of \emph{ASCA} data
\citep{ly01,bz03}. We  note again that our analysis  is an improvement
over the previous work because  we have calculated the excess variance
from light curves having nearly the same length.

The next step in  our work is to include the GIS  data in our analysis
to improve our  signal-to-noise and increase the number  of valid light
curves. We  will also  investigate the variance--mass  relationship in
different energy bands.

Most importantly, our future  work will involve simulations similar to
those used  in modeling AGN  power-spectra \cite[e.g.][]{mev03}.  This
will allow  us to account  for the sampling  of the light  curves and,
thereby,  permit a  robust  comparison with  a hypothesised  universal
power spectrum.  Moreover, we shall  quantify the extent to  which the
mean  excess  variance  of  each  object  differs  from  the  expected
variance.   These  residuals  can   then  be  examined  against  other
parameters,  such  as  mass-accretion  rate.   In  this  way,  we  can
investigate  the  extent  to  which  various  parameters,  other  than
black-hole mass, are associated with the X-ray variability.


\begin{theacknowledgments}

PMO acknowledges financial support from PPARC.  This research has made
use of  the TARTARUS database, which  is supported by  Jane Turner and
Kirpal Nandra under NASA grants NAG5-7385 and NAG5-7067.

\end{theacknowledgments}

\end{document}